\documentclass[12pt]{article}
 \usepackage{amsfonts}
 \usepackage{amssymb}
  \def\bsh{\backslash}

 \newfont{\bbbold}{msbm10}

\renewcommand{\thefootnote}{\fnsymbol{footnote}}
\renewcommand{\thanks}[1]{\footnote{#1}} 
\newcommand{\starttext}{
\setcounter{footnote}{0}
\renewcommand{\thefootnote}{\arabic{footnote}}}

\def\ba{\begin{eqnarray}}
\def\ea{\end{eqnarray}}
\def\no{\nonumber}

 \def\cA{{\cal A}}
 
 \def\cC{{\cal C}}

 \def\cN{{\cal N}}
 \def\cO{{\cal O}}
 \def\OO{{\cal O}}

 \newfont{\goth}{eufm10 scaled \magstep1}

 \def\gs{\mbox{\goth s}}
 
 \def\gu{\mbox{\goth u}}

\def\gsu{\gs\gu}
 
 \def\a{\alpha}
 \def\b{\beta}
 
 \def\d{\delta}
 \def\e{\epsilon}

 \def\L{\Lambda}

 \def\adt{\dot \alpha}
 \def\bdt{\dot \beta}

 \def\be{\begin{equation}}\def\ee{\end{equation}}
 \def\bea{\begin{eqnarray}}\def\eea{\end{eqnarray}}
 \def\ba{\begin{array}}\def\ea{\end{array}}

 \def\del{\partial}

 \def\str{\rm str}
 \def\xz{\times}

 \def\nab{\nabla}

 \def\del{\partial}

 \def\3dt{\dot{3}}
 
\def\half{{1\over2}}\def\qu{{1\over4}}
 \let\la=\label

 \let\bm=\bibitem{}

 \def\nn{\nonumber}
 \def\bd{\begin{document}}
 \def\ed{\end{document}}
 \def\bea{\begin{eqnarray}}\def\barr{\begin{array}}\def\earr{\end{array}}
 \def\eea{\end{eqnarray}}
 \def\ft#1#2{{\textstyle{{\scriptstyle #1}\over {\scriptstyle #2}}}}
 \def\fft#1#2{{#1 \over #2}}
 \newcommand{\eq}[1]{(\ref{#1})}
 \def\eqs#1#2{(\ref{#1}-\ref{#2})}
 \def\det{{\rm det\,}}
 \def\tr{{\rm tr}}\def\Tr{{\rm Tr}}
  \def\str{{\rm str}} \def\diag{{\rm diag}}
 \def\sdet{{\rm sdet}}

\newcommand{\ho}[1]{$\, ^{#1}$}
\newcommand{\hoch}[1]{$^{#1}$}

\begin{document}

\hfill{{UCLA/02/TEP/39}}
\\[-20pt]

\hfill{{KCL-MTH-03-01}}
\\[-20pt]

\hfill{{UUITP-09/02}}
\\[-20pt]

\hfill{hep-th/0301104}

\vspace{20pt}

\begin{center}


{\Large \bf Systematics of Quarter BPS operators in $\cN =4$
SYM}\footnote{Research supported in part by National Science
Foundation grants PHY-98-19686 and PHY-01-40151 and by the RTN European Program 
HPRN-CT-2000-00148.}


\vspace{40pt}

{ \bf E. D'Hoker\hoch1, P. Heslop\hoch{2,3} P. Howe\hoch4, and
A.V. Ryzhov\hoch1}\\[20pt]

{\sl \hoch1 Department of Physics and Astronomy, \\
 University of California, Los Angeles, CA  90095, USA}\\

{\sl \hoch2 II. Institut f\"ur Theoretische Physik 
der Universit\"at Hamburg}\\

{\sl \hoch3 Institut f\"ur Theoretische Physik, Universit\"at Leipzig}\\

{\sl \hoch4 Department of Mathematics, King's College, London,
U.K. }

\vspace{100pt}
 \newtheorem{definition}{Definition}
 \newtheorem{theorem}{Theorem}
 \newtheorem{proposition}{Proposition}

{\large\bf Abstract}

\end{center}

A systematic construction is presented of 1/4 BPS operators in
$\cN =4$ superconformal Yang-Mills theory, using either analytic
superspace methods or components. In the construction, the
operators of the classical theory annihilated by 4 out of 16
supercharges are arranged into two types. The first type consists
of those operators that contain 1/4 BPS operators in the full
quantum theory. The second type consists of descendants of
operators in long unprotected multiplets which develop anomalous
dimensions in the quantum theory. The 1/4 BPS operators of the
quantum theory are defined to be orthogonal to all the descendant
operators with the same classical quantum numbers. It is shown, to
order $g^2$, that these 1/4 BPS operators have protected
dimensions.

\hspace{5cm}

 \baselineskip=15pt \pagebreak \setcounter{page}{1}

\starttext
\section{Introduction}
\setcounter{equation}{0}

There are few things one can calculate exactly in the
four-dimensional superconformal Yang-Mills theories. One class of
such quantities consists of a set of correlation functions of the
Bogomolnyi Prasad Sommerfield (BPS) operators. In $\cN=4$
superconformal Yang-Mills, there are 1/2 BPS, 1/4 BPS and 1/8 BPS
operators, which are operators that are invariant under 8, 4 and 2
(out of 16) Poincar\'e supercharges respectively. Based on very
general arguments involving only the supersymmetry algebra
\cite{Dobrev85, Ferr, minwalla}, the anomalous dimension of any of
these operators vanishes identically in the full quantum theory.

\medskip

In the AdS/CFT correspondence, the local gauge invariant operators of
$\cN=4$ superconformal Yang-Mills are mapped to the physical
states of the Type IIB superstring on AdS$_5\times$S$^5$. (See
\cite{Malda, Gubs, witten} for the original papers and
\cite{magoo,Bigatti:1999dp,df02} for reviews.) The single trace
1/2 BPS operators (also referred to as chiral primary operators or
CPOs) play a special role as they are in one-to-one correspondence
with the short multiplets of supergravity and Kaluza-Klein states
with spins $\leq 2$. Driven by the success of this correspondence,
several authors have derived non-renormalization results for
various correlation functions of these operators. Results on the
perturbative non-renormalisation of 2- and 3-point functions were
derived in \cite{lmrs,FMMR,DFS} in components and in superspace in
\cite{pen99,HSW}; for further references on 3-point computations
see \cite{Bastianelli:1999vm}. An argument for the complete
non-renormalisation of the 3-point function of the supercurrent
multiplet based on anomalies was given in \cite{FMMR} and a
superspace version of this was presented in \cite{HSW}. An
argument for the (non-perturbative) non-renormalisation of all 2-
and 3-point functions of BPS operators based on an extra $U(1)_Y$
symmetry was given in \cite{int99} and this was verified using
analytic superspace methods in \cite{EHW} following on from  the
earlier work of \cite{HW}. Generalizations to $n$-point functions
were obtained for extremal \cite{DFMMR99B, Eden00, Bianchi99} and
near-extremal correlators \cite{Eden00A,
Erd99,DEFP,Arut99,Arutyunov:2001qw}.

\medskip

Other BPS operators are also important, both from the perspective
of superconformal Yang-Mills theory, and from that of the AdS/CFT
correspondence. The simplest generalization is to multi-trace 1/2
BPS operators \cite{Sk99}, for which non-renormalization results
are the same as for single trace 1/2 BPS operators; see also
\cite{Fer00}. Indeed, the arguments of \cite{int99} and\cite{EHW}
apply in this case too.

\medskip

A more delicate generalization is to the multi-trace scalar
operators obeying a 1/4 BPS shortening rule. A general group
theoretic classification of such operators in free field theory
was amongst the results derived in \cite{And99}. A more detailed
study carried out in \cite{Ryzhov:2001bp} revealed that in the
full quantum interacting theory, the true 1/4 BPS operators
involve admixtures of classical 1/4 BPS operators of \cite{And99}
with descendants of non-BPS operators that occur in long
supersymmetry multiplets. Using the operators obtained in
\cite{Ryzhov:2001bp}, 2- and 3-point functions involving 1/2 and
1/4 BPS operators were computed in \cite{dr,quarter}. In all cases
studied, these correlators were shown to be non-renormalized to
order $g^2$ as well as to be in accord with their large $N$, large
$g^2N$ limit accessible through the AdS/CFT correspondence. In the
framework of analytic superspace it turns out that the 1/4 BPS
operators are described as tensor superfields carrying
superindices. In \cite{HH} the 2- and 3-point functions of 1/4 BPS
operators were analysed in analytic superspace and shown to be
non-renormalised in a similar fashion to correlators of 1/2 BPS
operators. \footnote{In fact, this result can be extended to many
protected operators, including those which are in shortened series
A representations \cite{HH}.} Correlation functions with $n\geq 4$
operators are, in general, expected to receive quantum
corrections, just as the multipoint functions of 1/2 BPS operators
do \cite{df98}. (See also \cite{magoo, df02} for further
references.)

\medskip

However, an important puzzle arose in \cite{Ryzhov:2001bp}. It was
argued that the number of 1/4 BPS operators in the interacting
quantum theory for some representations was smaller than one would
have expected from counting the number of 1/4 BPS operators in the
classical theory. This would have required the presence of some
form of superconformal anomaly, which is not expected to be
present. Also, the procedure for constructing the candidate
operators used in \cite{Ryzhov:2001bp} was somewhat ad hoc and
difficult to generalize.

\medskip

In this note we use the machinery of (4,1,1) harmonic superspace
to describe 1/4 BPS operators. The use of extended supersymmetry
dramatically simplifies the counting and construction of scalar
composite operators in the $[q,p,q]$ representations of the
R-symmetry group $SU(4)$. We find that some operators were
overlooked in \cite{Ryzhov:2001bp}. Taking these operators into
account eliminates the mismatch between the number of 1/4 BPS
operators in the free and the interacting theory.

\medskip

The construction of all 1/4 BPS operators in the fully interacting
quantum theory is carried out as follows. First, in the classical
interacting theory, a basis is produced of all the scalar
operators in the representations of the R-symmetry group $SU(4)$
suitable for 1/4 BPS operators with Dynkin labels $[q,p,q]$,
$q\geq 1$ and with classical dimension $p+2q$. In the classical
interacting theory there are natural candidates for 1/4 BPS
operators as well as other operators which can be identified as
descendants. The basis may be regrouped into operators of these
two types. Harmonic superspace techniques reduce this construction
down to elementary group theory. The (4,1,1) superspace notation
also makes distinguishing candidate 1/4 BPS operators from
descendant operators simple.

\medskip

Already at the Born level, the two types of operators usually mix;
the overlaps between descendants and candidate BPS operators are
generally nonzero. In a given representation $[q,p,q]$ of $SU(4)$,
we identify the 1/4 BPS operators as those linear combinations of
candidate 1/4 BPS operators and descendants which at Born level
are orthogonal to all descendant operators in the same
representation of $SU(4)$. By construction therefore, the
admixture coefficients depend on $N$ but not on the coupling
$g^2$. A scalar composite operator in the $[q,p,q]$ of $SU(4)$
which is annihilated by 4 out of 16 (Poincar\'e) supercharges can
be either $(Q^2 \bar Q^2)$-descendants of long operators
or 1/4-BPS primaries.%
\footnote{
    It was shown in \cite{Ryzhov:2001bp} that
    these are the only possibilities by group theory.
} Thus, we argue that after subtracting off all the descendant
pieces we should be left with a protected operator. We then
proceed to calculate the two-point functions involving the 1/4 BPS
operators thus constructed. Remarkably, even though the
renormalization of long operators is governed by interactions, our
1/4 BPS operators constructed using Born level two-point functions
remain orthogonal to all the (descendants of) long operators at
order $g^2$. We find that any two point function involving a 1/4
BPS operator constructed this way receives no order $g^2$
corrections.

\medskip

In the following  we first review the 1/4 BPS operators from the
point of view of harmonic superspace and then discuss the
diagonalization of the operators corresponding to the same
representations that were discussed in \cite{Ryzhov:2001bp}.

\section{$\cN =4$ SYM in harmonic superspace}
\setcounter{equation}{0}

The leading component fields of the 1/4 BPS multiplets in $\cN =4$
SCFT are given by scalar fields which transform under the internal
symmetry group $SU(4)$  in representations with Dynkin labels of
the form  $[q,p,q]$. The complete supermultiplets can be very
simply described in harmonic superspace, and we briefly recall how
this construction works.

\medskip

For $\cN$ extended supersymmetry in four dimensions, $(\cN ,{\rm
p},{\rm q})$ harmonic superspace is obtained from ordinary
Minkowski superspace $M$ by the adjunction of a compact manifold
of the form $K \equiv H\bsh SU(\cN )$ where $H=S(U({\rm p})\xz
U(\cN -({\rm p}+{\rm q}))\xz U({\rm q}))$. This construction
allows one to construct p projections of the $\cN$ supercovariant
derivatives $D_{\a i}$ and q projections of their conjugates $\bar
D_{\adt}^i$ which mutually anticommute.  We can therefore define
generalized chiral or G-analytic superfields (G for Grassmann) in
such superspaces which are annihilated by these derivatives. Now
the superfields in harmonic superspace will also depend on the
coordinates of $K$, and as this space is a complex manifold, they
can be analytic in the usual sense (H-analytic) in their
dependence on these coordinates.  As the internal manifold is
compact H-analytic fields will have finite harmonic expansions.
The Lorentz scalar superfields which are both G-analytic and
H-analytic,  which we shall refer to as analytic, are the fields
we are interested in. They can be shown to carry short irreducible
unitary representations of the superconformal group (provided that
they transform under irreducible representations of $H$). For
original papers and detailed accounts of harmonic and analytic
superspaces, see for example \cite{gikos,
harthowe,superspace,DP99}.

\subsection{(4,1,1) Harmonic Superspace}

For the 1/4 BPS operators in $\cN=4$ the most appropriate harmonic
superspace has $(\cN,{\rm p},{\rm q})=(4,1,1)$. Since the analytic
fields in this space will be annihilated by one $D$ and one $\bar
D$ it follows that they will only depend at most on $3/4$ of the
odd coordinates of $M$.  Instead of working directly on the coset
defined by the isotropy group $H=S(U(1)\xz U(2)\xz U(1))$ we shall
follow the standard practice of working on the group $SU(4)$ which
amounts to the same thing provided that all the fields have their
dependence on $H$ fixed. We denote an element of $SU(4)$ by
$u_I{}^i$ and its inverse by $(u^{-1})_i{}^I$. The group $H$ is
taken to act on the capital index $I$ which we decompose as
$I=(1,r,4), \ r\in\{2,3\}$, while $SU(4)$ acts on the small
indices $i,j,\ldots$. Using $u$ and its inverse we can convert
$SU(4)$ indices into $H$ indices and vice versa. Thus we can
define
\be \ba{lll} D_{\a I}& \equiv u_I{}^i D_{\a i}&=(D_{\a 1},D_{\a
r}, D_{\a 4})
 \\
&&\\
 \bar D_{\adt}^I& \equiv \bar D_{\adt}^i (u^{-1})_i{}^I&= (\bar
D_{\adt}^1, \bar D_{\adt}^r, \bar D_{\adt}^4)
 \ea
\ee
Clearly, we have $\{D_{\a 1},\bar D_{\bdt}^4\}=0$. To
differentiate in the coset space directions we use the
right-invariant vector fields on $SU(4)$ which we denote by
$D_I{}^J$,  and which satisfy $\bar D^I{}_J=-D_J{}^I$ and
$D_I{}^I=0$.  These derivatives obey the Lie algebra relations of
$\gsu(4)$ and act on $u_K{}^k$ by
\be D_I{}^J u_K{}^k= \d_K{}^J u_I{}^k -\qu\d_I{}^J u_K{}^k \ee
The basic differential operators on $SU(4)$ can be divided into
three sets: the derivatives $(D_1{}^1,D_{r}{}^s,D_4{}^4)$
correspond to the isotropy group, the derivatives $(D_1{}^r,
D_1{}^4, D_r{}^4)$ can be thought of as essentially the components
of the $\bar\del$ operator on $K$ and the derivatives $(D_r{}^1,
D_4{}^1,  D_4{}^r)$ are the complex conjugates of these. Note that
the derivatives $(D_1{}^r, D_1{}^4, D_r{}^4)$ commute with $D_{\a
1}$ and $\bar D_{\adt}^4$. G-analytic fields are annihilated by
$D_{\a 1}$ and $\bar D_{\adt}^4$, H-analytic fields are
annihilated by $(D_1{}^r, D_1{}^4, D_r{}^4)$ and analytic fields
are annihilated by both of these sets of operators.

\medskip

The $\cN=4$ Yang-Mills theory is described in Minkowski superspace
by a scalar superfield $W_{ij}=-W_{ji}$ which transforms under the
six-dimensional representation of $SU(4)$ and also under the
adjoint representation of the gauge group which we take to be
$SU(N)$. It is real in the sense that $\bar W^{ij}=\half \e^{ijkl}
W_{kl}$. This superfield satisfies the constraints
\bea \nab_{\a i} W_{jk}&=&\e_{ijkl} \L_{\a}{}^l
\no \\
\bar \nab_{\adt}^i W_{jk}&=& 2\d^i_{[j}\bar\L_{\adt k]} \eea
where $\L$ is a superfield whose leading component is the spinor
field of the multiplet.and where $\nab_{\a i}$ is a spinorial
derivative which is covariant with respect to the gauge group.
Using the superspace Bianchi identities one can easily show that
the only other independent spacetime component of $W$ is the
spacetime Yang-Mills field strength and that all of the component
fields satisfy their equations of motion.

\medskip

In $(4,1,1)$ superspace we can define the superfield $W_{1r}\equiv
u_1{}^i u_r{}^j W_{ij}$. Using the properties outlined above one
can easily show that
\be
\nab_{\a 1} W_{1r}=\bar\nab_{\adt}^4 W_{1r}=0
\label{eq:delta W = 0}
\ee
and that the derivatives  $(D_1{}^r, D_1{}^4, D_r{}^4)$  all
annihilate $W_{1r}$, so that $W_{1r}$ is a covariantly analytic
field on $(4,1,1)$ harmonic superspace.  However, if we consider
gauge-invariant products of $W$'s, i.e. traces or multi-traces,
the resulting objects will be analytic superfields; they will be
annihilated by $D_{\a 1}$ and $\bar D_{\adt}^4$ rather than the
gauge-covariant versions. These are the superfields which we shall
use to construct the 1/4 BPS states. To make the formulae less
cluttered we shall abbreviate%
\footnote{
    So the index $r$ takes the values $r= 2, 3$ for $W_{1r}$,
    and $r = 1,2$ for $W_r$.
    }
$W_{1r}$ to $W_{r-1}$ and define
$W^r\equiv \e^{rs} W_s$.

\subsection{Quarter BPS Operators}

These superfields are easy to describe. The superfield
corresponding to the representation $[q,p,q]$ contains $p+2q$
powers of $W$ in the representation $p$ of $SU(2)$ (this is the
$SU(2)$ in the isotropy group), i.e. it has $p$ symmetrized
$SU(2)$ indices. If $q=0$ the single trace operators are the
chiral primaries which are 1/2 BPS. These operators we will refer
to as CPOs and denote by $A_p$,
\be A_{r_1\ldots r_p}\equiv \tr (W_{(r_1}\ldots W_{r_p)}) \ee
The lowest CPO is the stress-tensor multiplet $T_{rs}=A_{rs}$. We
can obtain further 1/2 BPS operators by taking products of CPOs
and  symmetrizing on all of the $SU(2)$ indices. The 1/4 BPS
operators (for $q > 0$) fall into two classes. There are operators
that can be constructed as products of the CPOs with at least one
pair of contracted indices, for example \be T_{rs} T^{rs} ,\quad
A_{rst} A^{st} ,\quad A_{rst} A^t{}_{uvw} ,\quad \mbox{and so on.}
\ee These operators have no commutators in their definition, and
so are the candidate 1/4 BPS operators, up to subtleties which we
shall come to in due course. Operators in the other class have at
least one single-trace factor in which the indices of two or more
pairs of $W$'s are contracted, as in \be \tr W^2 W^2 ,\quad
A_{rst} \; \tr W^2 W^2 ,\quad \tr W_r W_s W^2 W^2 ,\quad
\mbox{etc.} \ee where
 \be
W^2\equiv W_r W^r=\e^{rs} W_r W_s =\half \, \e^{rs} [W_r,W_s]
 \ee
These operators are descendants; the superspace Bianchi identities
imply that
\be \e^{\a\b}\nab_{\a i} \nab_{\b j} \bar W^{kl}= 2\d_j{}^{[k}
[W_{im}, \bar W^{l]m}] \ee
>From this formula and its conjugate one can see that $W^2$ can be
written as
 \be
 W^2= (\nab_1)^2 \bar W^{14}=-(\bar\nab^4)^2 W_{14}
 \label{eq:delta = W2}
 \ee
where $(\nab_1)^2\equiv  1/2 \e^{\a\b}\nab_{\a 1}\nab _{\b 1}$,
$(\bar\nab^4)^2\equiv -1/2 \e^{\adt\bdt}
\bar\nab_{\adt}^4\bar\nab_{\bdt}^4$ and where $W_{14}=u_1{}^i
u_4{}^j W_{ij}$. Given a product of $W_r$'s containing a factor of
$W^2$, therefore, the latter can be written in terms of
derivatives as above and the derivatives can be taken to act on
the whole expression with $W^2$ replaced by either $W_{14}$ or its
conjugate. This follows by G-analyticity. Indeed, if there are two
factors of $W^2$ in an operator then all four derivatives can be
brought outside. This is because
 \be
 \nab_{\a 1} W_{14}=0\qquad {\rm and}\qquad
 \bar\nab_{\adt}^4 \bar W^{14}=0
 \label{eq:delta4 = 0}
 \ee
In fact, the descendant 1/4 BPS operators always have at least two
factors of $W^2$ so that they can be written explicitly as
derivatives of long operators by these means. However, we are not
quite finished yet because the ancestor operators will not be
H-analytic on $(4,1,1)$ harmonic superspace as they stand. This
can be remedied by noting that
\be
(\nab_1)^2(\bar \nab^4)^2 \bar W^{1r}=[W^2,W^r]
\label{eq:delta^4 W}
\ee
with the aid of which we can write, for example, $\tr(W^2 W^2)$ as
\bea
 \label{eq:konishi descendant}
 \tr(W^2W^2)&=&-{1\over3}(\nab_1)^2(\bar \nab^4)^2 \left( \tr(W_{14}\bar
W^{14})+\tr(W_{1r}\bar W^{1r})\right)\nn\\
&=&-{1\over12} (\nab_1)^2(\bar \nab^4)^2 \tr(W_{ij}\bar
W^{ij})\nn\\
&=&-{1\over12} (D_1)^2(\bar D^4)^2 \tr(W_{ij}\bar W^{ij}) \eea
where $D_{\a 1}D_{\b 1}=\e_{\a\b}(D_1)^2$, and $\bar D_{\adt}^
4\bar D_{\bdt}^4=-\e_{\adt\bdt}(\bar D^4)^2$. Hence we see that
$\tr(W^2 W^2)$ is a descendant of the Konishi operator
$K\equiv\tr(W_{ij}\bar W^{ij})$. In general, one can use
(\ref{eq:delta W = 0}), (\ref{eq:delta = W2}),
(\ref{eq:delta4 = 0}) and (\ref{eq:delta^4 W}) to find
\bea
 \label{eq:konishi-like general descendant}
{1\over 2} (\nab_1)^2(\bar \nab^4)^2
\; W_{ij} A \bar W^{ij}
=
W_r A [W^2, W^r] + [W^2, W^r] A W_r - 2 W^2 A W^2
\eea
provided $A$ involves only the $W_r$.
And since we are dealing with gauge invariant operators,
we can replace the $\nab$ by $D$.
We note for future use that each of the descendants that we
consider below can be written as an ancestor superfield acted on
by the differential operator $(D_1)^2(\bar D^4)^2$.

On the other hand, the CPOs themselves cannot be obtained by
differentiation from other operators and so the candidate 1/4 BPS
operators cannot be (entirely) descendants. An operator
annihilated by $D_1$ and $\bar D^4$ can be either a
$(D_1)^2(\bar D^4)^2$ descendant of a long primary; or a
$(D_1)^2$ or $(\bar D^4)^2$ descendant of a 1/8 BPS primary;
or a 1/4 BPS primary. In \cite{Ryzhov:2001bp} it was shown that a
$[q,p,q]$ scalar composite operator can not be a descendant of a
1/8 BPS primary. Therefore, we argue that after subtracting off
all the descendant pieces from candidate BPS operators, we should
be left with a 1/4 BPS primary; it simply can not be anything
else!

\subsection{Examples of systematic description}
\label{section:systematic description}

We begin by outlining a few rules that determine which tensor
structures are permitted.
First we observe that, since contractions are made using the
antisymmetric tensor $\e^{rs}$ while the tensors $A_{rs ... t}$
are symmetric, contractions within the same $A$ give zero,
\begin{eqnarray}
A^r{}_{r s ... t} = 0 ,
\end{eqnarray}
so we can only contract indices in different $A$'s.

Next, consider $T^2$. Since $T_{rs}$ transforms under the
3-dimensional representation of $SU(2)$ it follows that the
product of two $T$s will decompose into the five and
one-dimensional representations. The former corresponds to
symmetrisation on all four indices, i.e. $T_{(rs}T_{uv)}$, while
for a single contraction we have
 \be
 \label{eq:T-T:contract:general}
 T_{r t} T_s{}^t=-T_{s t} T_r{}^t=\half\e_{rs}T_{uv}T^{uv}
 \ee
Similarly, the product of three $T$s contains only the seven-and
three-dimensional representations, so that, for example
 \be
 \label{eq:T-T:contract}
 T_r{}^s T_s{}^t T_t{}^r=0
 \ee

One can look at contractions of other $A$'s in a similar fashion.
For example, $(A_3)^2$ contains only the seven- and
three-dimensional representations of $SU(2)$ corresponding to
tensors obtained by symmetrising on six or two indices with zero
or two contractions respectively. So
 \be
 \label{eq:A3-A3:contract}
 A^{rst} A_{rst}=A_{(rs}{}^v A_{tu)v}=0
 \ee
while
 \be
 A_{r}{}^{tu}A_{s tu}=A_{s}{}^{tu}A_{r tu}
 \ee
or, equivalently,
\begin{eqnarray}
\label{eq:A3-A3:contract:general} 3 A_{rst} A^t{}_{uv} &=& \e_{r
u} ( A_{s t w} A^{t w}{}_v ) + \e_{r v} ( A_{s t w} A^{t w}{}_u )
+ \e_{s v} ( A_{r t w} A^{t w}{}_u ) + \e_{s u} ( A_{r t w} A^{t
w}{}_v ) . \hspace{3em}
\end{eqnarray}

These equations  generalise in a straightforward manner. Whenever
we contract an odd number of indices in two $A$s of the same
length and symmetrize on the
remaining indices, the resulting tensor vanishes.%
\footnote{
    In general there will be more than one nonvanishing structure.
    For instance, both
    $A_{(rs}{}^{vw} A_{tu)vw}$ and
    $A^{rstu} A_{rstu}$
    are independent nonvanishing tensors.
} If $A$s of different length are contracted (as in $A_{rs}{}^t
T_{tu}$), there is no such restriction.

We shall now discuss some explicit examples. We shall use the
convention that uncontracted $SU(2)$ indices are understood to be
totally symmetrized.

\subsubsection*{The Representation $[1,p,1]$}

Such operators have to have $(p+2)$ $W_r$'s and only one
contraction. There are no single trace operators in this class
because
 \be
 \tr(W_{r_1}\ldots W_{r_p}W^2)=0
 \ee
Hence these operators can only be constructed by contracting CPOs.
They are all protected. This can also be seen from representation
theory because there are no long representations which contain
these representations \cite{Dolan:2002zh}. This result also shows
that any single-trace factor in an operator must have at least two
contractions.

Here we list the lowest dimensional examples of $[1,p,1]$
representations (for $p \le 5$). For [1,1,1] and [1,2,1] we can
not construct any nonvanishing tensors of this form. For [1,3,1],
there is one possible operator,
 \bea
 \label{eq:ops:131}
 \cO&=& A_{rst} T^{t}{}_{u}
 \eea
Similarly, for [1,4,1] the only possible operator is
 \bea
 \label{eq:ops:141}
 \cO&=& A_{rstu} T^{u}{}_{v}
 \eea
Higher representations offer more choices, and already in the
[1,5,1] we find
 \bea
 \label{eq:ops:151}
 \cO_1&=& A_{rstuv} T^{v}{}_{w}
\no \\
 \cO_2&=& A_{rstv} A^{v}{}_{uw}
\no \\
 \cO_3&=& T_{rs} T_{tv} A^{v}{}_{uw}
 \eea
All of these operators have protected two-point functions, as we
will explicitly verify in Section \ref{section:1p1:details}.

\subsubsection*{The Representation [2,0,2]}

This operator is realized as an $SU(2)$ scalar in $(4,1,1)$
harmonic superspace. There are just two possibilities
 \bea
 \label{eq:ops:202}
 \cO_1&=& T_{rs} T^{rs}
\no \\
 \cO_2 &=& \tr(W^2 W^2)
 \eea
Using the rules outlined in the beginning of Section
\ref{section:systematic description}, we see that $\cO_1$ is the
only multiple trace operator one can construct with two pairs of
contracted indices, and there is also no other choice for the
single trace operator but $\cO_2$. $\cO_1$ is a candidate 1/4 BPS
operator while $\cO_2$ is a descendant; as seen in
(\ref{eq:konishi descendant}), it is a descendant of the Konishi
operator.

\subsubsection*{The Representation [2,1,2]}

This has 5 fields and forms an $SU(2)$ doublet. There are again
only two possibilities
 \bea
 \cO_1&=& A_{rst} T^{st}
\no \\
 \cO_2&=& \tr(W_r W^2 W^2)
 \eea
This case is completely parallel to the [2,0,2] representation.

The operator $\cO_2$ can be written in the form
 \be
 \cO_2=-{1\over16} (D_1)^2(\bar D^4)^2 \,\tr(W_r W_{ij} \bar W^{ij})
 \ee
Note that the ancestor here is defined on (4,1,1) harmonic
superspace; it is not G-analytic but it is H-analytic. One can
easily remove the harmonic variables to obtain the corresponding
superfield on ordinary superspace. In this case it is
$\tr(W_{ij}W_{kl}\bar W^{kl})$.

\subsubsection*{The Representation [2,2,2]}

This has 6 fields and transforms as a triplet under $SU(2)$.
Multiple trace operators are constructed in the following way. We
can partition the set of six fields as 6 = 4 + 2, 6 = 3 + 3, or 6
= 2 + 2 + 2. Two pairs of indices are contracted, and two
remaining indices are symmetrized. The possibilities are
 \bea
 \label{eq:222:fund}
 \cO_1&=& A_{rstu} T^{tu}
\no \\
 \cO_2&=& A_{r}{}^{tu} A_{stu}
\no \\
 \cO_3&=& T_{rs} T_{tu} T^{tu}
\no \\
 \cO_4&=& \tr(W_r W_s W^2 W^2)
\no \\
 \cO_5&=& \tr(W_r W^2 W_s W^2)
\no \\
 \cO_6&=& \tr (W^2 W^2) T_{rs}
 \eea
The first three are candidate 1/4 BPS operators while the last
three are descendants. For the partitions 6 = 4 + 2 and 6 = 3 + 3,
these are the only choices because contractions within the same
$A$ give zero. For the partition 6 = 2 + 2 + 2, equation
(\ref{eq:T-T:contract}) relates any other triple-trace [2,2,2]
operator to $\cO_3$.

The operator $\cO_6$ is a descendant of the product of the Konishi
operator and the supercurrent, while $\cO_4$ and $\cO_5$ are
descendants of the operators
 \bea
 \cA_1&=&\tr(W_r W_s W_{ij}\bar W^{ij})
 \no\\
 \cA_2&=&\tr(W_r W_{ij} W_s\bar W^{ij})
 \eea
A short calculation yields
 \bea
 \cO_4&=&{1\over40}(D_1)^2(\bar D^4)^2(\cA_2-3 \cA_1)
 \no\\
 \cO_5&=&{1\over20}(D_1)^2(\bar D^4)^2(\cA_1- 2 \cA_2)
 \eea
In terms of ordinary superfields, both $\cA_1$ and $\cA_2$ are in
the $[0,2,0]$ representation of $SU(4)$.

\subsubsection*{The Representation [2,3,2]}

The possibilities are
 \bea
 \label{ops:232}
 \cO_1&=& A_{rstuv} T^{uv}
\no \\
 \cO_2&=& A_{rs}{}^{uv} A_{tuv}
\no \\
 \cO_3&=& A_{rst} T_{uv} T^{uv}
\no \\
 \cO_4&=& A_r{}^{uv} (T^2)_{stuv}
\no \\
 \cO_5&=& \tr(W_r W_s W_t W^2 W^2)
\no \\
 \cO_6&=& \tr(W_r W_s W^2 W_t W^2)
\no \\
 \cO_7&=& T_{rs}\tr(W_t W^2 W^2)
\no \\
 \cO_8&=& A_{rst}\tr(W^2W^2)
 \eea
The first four are candidate 1/4 BPS operators while the second
four are descendants.

The last of these is a descendant of a product of $A_3$
and the Konishi operator, while the ancestor of $\cO_7$ is a
product of $T$ and $\tr(W_r W_{ij}\bar W^{ij})$. For the other two
descendants we have
 \bea
 \cO_5&=&{1\over24}(D_1)^2(\bar D^4)^2(\cA_2-2\cA_1)
 \no\\
 \cO_6&=&{1\over24}(D_1)^2(\bar D^4)^2(\cA_1-2\cA_2)
 \eea
where
 \bea
 \cA_1&=&\tr(W_rW_sW_t W_{ij}\bar W^{ij})
 \no\\
 \cA_2&=&\tr(W_rW_s W_{ij}W_t\bar W^{ij})
 \eea

\subsubsection*{The Representation [3,0,3]}

There is only one possibility: \bea \cO &=& \tr(W^2 W^2 W^2) \eea
This operator is a descendant. There are no candidate 1/4 BPS
operators in this case. As we saw in (\ref{eq:T-T:contract}) and
(\ref{eq:A3-A3:contract}), the operators $A_{rst} A^{rst}$ and
$T_{rs} T_t{}^r T^{st}$ vanish identically. Explicitly, this
operator can be written as
 \be
 \cO=-{1\over8}(D_1)^2(\bar D^4)^2\tr(W^2 W_{ij}\bar W^{ij})
 \ee

\subsubsection*{The Representation [3,1,3]}

This example again has seven fields but the representation of
$SU(2)$ is the doublet. The operators are
 \bea
 \label{ops:313}
 \cO_1&=& A_r{}^{stu} A_{stu}
\no \\
 \cO_2&=& (T^2)_{rstu} A^{stu}
\no \\
 \cO_3&=& \tr(W_r W^2 W^2 W^2)
\no \\
 \cO_4&=& \tr(W_r W^2 W_s W^2 W^s - W_r W_s W^2 W^s W^2)
\no \\
 \cO_5&=& T_r{}^s \tr(W_s W^2 W^2)
 \eea
so there are 3 descendants in this case. We have symmetrized
$\cO_4$ so that $\cO_4^\dagger = + (\cO_4)^*$. This symmetry
amounts to charge conjugation on the fields $X$ in the adjoint
representation of the gauge group and the ${\bf 6}$ of $SU(4)$.
Its effect on the $\cN=1$ superfield formulation is to map $z_j
\to z_j ^t$.

The last operator is again a descendant of a product of operators
that we have discussed previously. For the other two we have
 \bea
 \cO_3&=&{1\over30}(D_1)^2(\bar D^4)^2(\cA_2-5\cA_1)
 \no\\
 \cO_4&=&{1\over6} (D_1)^2(\bar D^4)^2(\cA_1+\cA_2)
 \eea
where
 \bea
 \cA_1&=&\tr(W_r W^2 W_{ij}\bar W^{ij})
 \no\\
 \cA_2&=&\tr(W_r W_s W_{ij} W^s\bar W^{ij})
 \eea

\subsection{Multiplicity of quarter BPS operators}

The (classical) quarter BPS operators in the $SU(4)$
representation are built from single trace quarter (and half) BPS
operators. These have the form \be
 \tr (W_{(r_1}\dots W_{r_q)}
(W^2)^p) \ee for operators in the $[p,q,p]$ $SU(4)$
representation, but the order of the $2p+q$ operators inside the
trace is arbitrary.

To find the number of different single trace operators in this
representation, $N_{pq}$, consider the reducible operator
\be X_Q:=\tr(W_{r_1}\dots W_{r_Q}), \ee
where the $SU(2)$ indices are no longer taken to be symmetrised.
This is in a reducible representation of $SU(2)$, and contains all
single trace scalar composite operators of dimension $Q$.
So one obtains the number of
operators in each representation by
expanding this operator as a sum of irreducible representations.
For example to find all single trace
operators of dimension 4 consider $X_4$. This has 6 components
(given by (1111), (1112), (1122), (1212), (2221), (2222)
where $(r_1 r_2 r_3 r_4)$
is short hand for $\tr(W_{r_1}\dots W_{r_4})$.) In
terms of irreducible $SU(2)$ representations it splits as $6=5+1$.
In terms of $SU(4)$ representations the 5 corresponds to $[0,4,0]$
and the 1 corresponds to $[2,0,2]$, and so we find that there is
only one operator in each of these two representations.

More generally, to split $X_Q$ into irreducibles, consider the
components of $X_Q$. Let $c(Q,p)$ denote the number of components
of $X_Q$ with $p$ 1's and $Q-p$ 2's, i.e. the number of ways to
arrange a total of $Q$ objects with $p$ of one type and $Q-p$ of
another type up to circular permutations.

Then $X_Q$ splits into the following irreducible representations:
\be \sum_{p=0}^{\lfloor Q/2 \rfloor} \left(c(Q,p) - c(Q,p-1)
\right) \ [p,q,p] \ee where $q=Q-2p$ and where $\lfloor x \rfloor$
denotes the largest integer less than or equal to $x$. So the
number of single trace operators in the $[p,q,p]$ representation
is \be N_{pq}=c(Q,p)-c(Q,p-1).\la{mult} \ee In general the formula
for $c(Q,p)$ is quite complicated, but in certain cases it
simplifies. For example
\be c(Q,0)=1
,\quad
c(Q,1)=1
,\quad
c(Q,2)=\lfloor {Q /2} \rfloor
,
\ee
and if $Q$ and $p$ are co-prime
then \be
 N_{pq}= {1 \over Q} {Q \choose p}.
\ee

As an example, consider dimension 6 operators: $X_6$ has 14
components and $c(6,p)$ is given by:
\be c(6,0)=c(6,1)=1
,\quad
c(6,2)=3
,\quad
c(6,3)=4.
\ee
Then~\eq{mult} gives
\be N_{06}=1
,\quad
N_{14}=0
,\quad
N_{22} = 2
,\quad
N_{30}=1
,
\ee
reproducing the
correct numbers of single-trace operators discussed above (in
particular there are two operators in the $[2,2,2]$ representation
and 1 in the $[3,0,3]$.)

Since multiple trace operators can be obtained by multiplying
together single trace operators, to find the number of multi-trace
operators in a given representation one just has to consider all
possible ways of obtaining the representation in question from
tensor products of other representations and use the formula for
single-trace operators.

\subsection{Relationship between $\cN=4$ and $\cN=1$ superfields}

The map between quarter BPS operators in the $\cN=1$ formalism and
those in $(4,1,1)$ analytic superspace is straightforward. In the
$\cN=1$ formalism the quarter BPS operators are given by
\be \left[ (z^{2c})^p (z_d)^q \right] \ee
where $[\dots ]$ denote gauge invariant combinations,
$(X_a)^p\equiv X_{(a_1}\dots X_{a_p)}$ and $z^{2c}\equiv z_a z_b
\e^{abc}$. Here the $a,b,\dots = (1,2,3)$ are $SU(3)$ indices.
These operators have highest weight state given by
\be \left[ (z_{1} z_{2}-z_{2} z_{1} )^p (z_1)^q \right]. \ee
In $(4,1,1)$ harmonic superspace on the other hand, this object is
given by
\be \left[ (W^2)^p  (W_r)^q  \right]. \ee
If we relabel the $SU(2)$ indices $r,s,\dots =1,2$ then this
operator has highest weight state
\be \left[ (W_1 W_2-W_2 W_1)^p  (W_1)^q  \right]. \ee
The correspondence between the $N=1$ operators and the harmonic
superspace operators is now clear, one simply replaces $W$ with
$z$ to obtain the highest weight states of each.

\section{Explicit Computations}
\setcounter{equation}{0}

In this section we will explicitly calculate two point functions
of the above operators. We will work with the lowest components of
superfields, the $z_i^a$ and $\bar z_i^a$. (Here, $i=1, ... , 3$,
and $a$ labels the adjoint representation of the gauge group
$SU(N)$.) We list operators in a given irrep of the R-symmetry
group $SU(4)$ (some of them were missed in \cite{Ryzhov:2001bp}),
and for the descendant operators write out the corresponding
Konishi-like long operator they come from. Then we look at Born
level and order $g^2$ contributions to the two point functions of
the highest weight state operators. (Most of them were calculated
in \cite{Ryzhov:2001bp}).

In each representation we will have the descendant operators $L_i$
and ``candidate 1/4-BPS'' operators $\OO$. We will compute the
order $g^0$ two point functions $\langle O L_i^\dagger
\rangle_{\rm Born}$ and $\langle L_i L_j^\dagger \rangle_{\rm
Born}$. Then we will consider operators \bea \tilde{\OO} \equiv
\OO - \langle O L_i^\dagger \rangle_{\rm Born} \biggl (\langle L
L^\dagger \rangle_{\rm Born}^{-1} \biggr )^{ij} L_j \eea By
construction, they are orthogonal to all the $L_i$ at Born level,
$\langle \tilde{\OO} L_i^\dagger \rangle_{\rm Born} = 0$. Then we
will show that these operators $\tilde \OO$ have protected two
point functions at order $g^2$, $\langle \tilde{\OO} L_i^\dagger
\rangle_{g^2} = 0$ and $\langle \tilde{\OO}' \tilde \OO^\dagger
\rangle_{g^2} = 0$ for all such operators $\tilde \OO, \tilde
\OO'$. The claim is that these operators $\tilde{\OO}$ are
1/4-BPS.

The basis of operators we will choose is slightly different from
the one used in \cite{Ryzhov:2001bp}. The operators introduced in
the preceding section are more natural and intuitive.

\subsubsection*{The representation $[1,p,1]$}
\label{section:1p1:details}

$ \bullet $ There is only one operator in the representation
$[1,3,1]$ whose highest $SU(4)$ weight state is \bea \OO &\equiv&
\tr z_1 z_1 \; \tr z_1 z_1 z_2 - \tr z_1 z_2 \; \tr z_1 z_1 z_1
\eea while acting on $\OO$ once with an $SU(4)$ ladder operator
gives \bea \OO' &\equiv& 2 \, \tr z_1 z_1 \; \tr z_1 z_2 z_2 - \tr
z_1 z_2 \; \tr z_1 z_1 z_2 - \tr z_2 z_2 \; \tr z_1 z_1 z_1 \eea
This operator has the same weight as the [2,1,2] operators (but is
of course orthogonal to them). The Born and order $g^2$ overlaps
are \bea \label{eq:131:born} \langle \OO' \OO'{}^\dagger
\rangle_{\rm Born} &=& {15 \over 32} N (N^2 - 1) (N^2 - 4) , \quad
\langle \OO' \OO'{}^\dagger \rangle_{g^2} = 0 .
\end{eqnarray}
So indeed it is a 1/4-BPS operator.

\medskip

$\bullet$ There is only one operator in the representation
$[1,4,1]$.  The highest $SU(4)$ weight state operator is \bea \OO
&\equiv& \tr z_1 z_1 \; \tr z_1 z_1 z_1 z_2 - \tr z_1 z_2 \; \tr
z_1 z_1 z_1 z_1 \eea while acting on $\OO$ once with an $SU(4)$
ladder operator gives \bea \label{eq:141:ops} \OO' &\equiv& 2 \,
\tr z_1 z_1 \; \tr z_1 z_1 z_2 z_2 + \tr z_1 z_1 \; \tr z_1 z_2
z_1 z_2 \nonumber\\&& - 2 \, \tr z_1 z_2 \; \tr z_1 z_1 z_1 z_2 -
\tr z_2 z_2 \; \tr z_1 z_1 z_1 z_1 \eea This operator has the same
weight as the [2,2,2] operators (but is of course orthogonal to
them). The Born and order $g^2$ overlaps are \bea
\label{eq:141:born} \langle \OO' \OO'{}^\dagger \rangle_{\rm Born}
&=& {3 \over 8} (N^2 - 1) (N^2 - 4) (N^2 - 9) , \quad \langle \OO'
\OO'{}^\dagger \rangle_{g^2} = 0 . \eea So indeed it is a 1/4-BPS
operator.

\medskip

$\bullet$ Finally, there are 3 operators in the representation
$[1,5,1]$. Their highest $SU(4)$ weight state operators are \bea
\OO_1 &\equiv& \tr z_1 z_1 \; \tr z_1 z_1 z_1 z_1 z_2 - \tr z_1
z_2 \; \tr z_1 z_1 z_1 z_1 z_1
\\
\OO_2 &\equiv& \tr z_1 z_1 z_1 \; \tr z_1 z_1 z_1 z_2 - \tr z_1
z_1 z_2 \; \tr z_1 z_1 z_1 z_1
\\
\OO_3 &\equiv& \tr z_1 z_1 \left( \tr z_1 z_1 \; \tr z_1 z_1 z_2 -
\tr z_1 z_2 \; \tr z_1 z_1 z_1 \right) \eea while acting on $\OO$
once with an $SU(4)$ ladder operator gives \bea \OO_1' &\equiv& 2
\, \tr z_1 z_1 \; \tr z_1 z_1 z_1 z_2 z_2 + 2 \, \tr z_1 z_1 \;
\tr z_1 z_1 z_2 z_1 z_2 \nonumber\\&& - 3 \, \tr z_1 z_2 \; \tr
z_1 z_1 z_1 z_1 z_2 - \tr z_2 z_2 \; \tr z_1 z_1 z_1 z_1 z_1
\\
\OO_2' &\equiv& 2 \, \tr z_1 z_1 z_1 \; \tr z_1 z_1 z_2 z_2 + \tr
z_1 z_1 z_1 \; \tr z_1 z_2 z_1 z_2 \nonumber\\&& - \tr z_1 z_1 z_2
\; \tr z_1 z_1 z_1 z_2 - 2 \, \tr z_1 z_2 z_2 \; \tr z_1 z_1 z_1
z_1
\\
\OO_3' &\equiv& 2 \, \tr z_1 z_1 \; \tr z_1 z_1 \; \tr z_1 z_2 z_2
+ \tr z_1 z_1 \; \tr z_1 z_2 \; \tr z_1 z_1 z_2 \nonumber\\&& - 2
\, \tr z_1 z_2 \; \tr z_1 z_2 \; \tr z_1 z_1 z_1 - \tr z_2 z_2 \;
\tr z_1 z_1 \; \tr z_1 z_1 z_1 \eea These operators have the same
weight as the [2,3,2] operators (but are of course orthogonal to
them). The Born and order $g^2$ overlaps are \bea
\label{eq:151:born} \langle \OO_i' \OO_j'{}^\dagger \rangle_{\rm
Born} &=& \mbox{$ {35 (N^2 - 1) (N^2 - 4) \over 128 N} $}
\nonumber\\&& \hspace{-3em}\times \left(\matrix{ N^4 - 10 N^2 + 72
& -11 N^2 + 36     & 6 N (N^2 - 2)      \cr
                  & N^4 - 4 N^2 + 18 & - 2 N ( 2 N^2 + 3) \cr
                  &                & 2 N^2 (N^2 + 5)
}\right) , \quad
\\
\langle \OO_i' \OO_j'{}^\dagger \rangle_{g^2} &=& 0 . \eea So
indeed they all are 1/4-BPS operators.

\subsubsection*{The Representation [2,0,2]}
\label{section:202:details}

The operators corresponding to (\ref{eq:ops:202}) are \bea
\label{ops:202} \OO_1 &=& 2 \left( \tr z_1 z_1 ~ \tr z_2 z_2 - \tr
z_1 z_2 ~ \tr z_1 z_2 \right)
\no \\
\OO_2 &=& \tr z_1 z_1 z_2 z_2 - \tr z_1 z_2 z_1 z_2 \eea The
single trace operator $\OO_2$ is a descendant of the Konishi
scalar, \bea (\bar Q_{\bar \zeta})^2 \; \tr z_j \bar z^j &=& \bar
Q_{\bar \zeta} \tr z_j \sqrt{2} \bar\zeta \bar\psi^j =
6 i \sqrt{2} (\bar\zeta \bar\zeta) \tr [z_1,z_2] z_3 ,
\no \\
(Q_{\zeta_3})^2 (\bar Q_{\bar \zeta})^2 \; \tr z_j \bar z^j &=& -
12 i (\bar\zeta \bar\zeta) Q_{\zeta_3} \tr [z_1,z_2] \zeta_3
\lambda \no \\ & = & 24 (\bar\zeta \bar\zeta) (\zeta_3 \zeta_3)
\tr [z_1,z_2]^2
\nonumber\\
&=& - 48 (\bar\zeta \bar\zeta) (\zeta_3 \zeta_3) \; \OO_2 ;
\label{konishi:descendant} \eea or
\begin{eqnarray}
\OO_2 \sim (Q^2 \bar Q^2 ) \; \tr z_j \bar z^j .
\label{eq:konishi:202} \eea
for short.%
\footnote{
    We will not write out the indices of the
    supercharges or proportionality constant explicitly from now on.
    The supercharges will be always the same
    as in (\ref{konishi:descendant}),
    and keeping track of all the factors like 48 or
    $(\bar\zeta \bar\zeta) (\zeta_3 \zeta_3)$
    would only clutter the notation.
} On the other hand, the operator orthogonal to $\OO_2$
at Born level
\begin{eqnarray}
\tilde \OO_1 = \OO_1 - {4 \over N} \OO_2
\label{eq:bps:202}
\eea
stays orthogonal to $\OO_2$,
$\langle \tilde \OO_1(x) \bar \OO_2(y) \rangle = 0$;
and has a two-point function
$\langle \tilde \OO_1(x) \bar {\tilde\OO}_1(y) \rangle = \langle
\tilde \OO_1(x) \bar {\tilde\OO}_1(y) \rangle_{\rm Born}$
protected at order $g^2$.

\subsubsection*{The Representation [2,1,2]}
\label{section:212:details}

In this representation we again have only two operators \bea \OO_1
&=& \tr z_1 z_1 z_1 ~ \tr z_2 z_2 - 2 \tr z_1 z_2 ~ \tr z_1 z_2
z_1 + \tr z_1 z_2 z_2 ~ \tr z_1 z_1
\no \\
\OO_2 &=& \tr z_1 z_1 z_1 z_2 z_2 - \tr z_1 z_1 z_2 z_1 z_2 \eea
and the single trace operator is again a descendant, \bea \OO_2
\sim (Q^2 \bar Q^2 )
\; \tr \left[ z_1 z_j \bar z^j + z_1 \bar z^j z_j \right]
\label{eq:konishi:212}
\eea
The operator orthogonal to $\OO_2$ at Born level
\bea \tilde \OO_1 = \OO_1 - {6 \over N} \OO_2
\label{eq:bps:212}
\eea
satisfies $\langle \tilde \OO_1(x) \bar \OO_2(y) \rangle =
0$, $\langle \tilde \OO_1(x) \bar {\tilde\OO}_1(y) \rangle =
\langle \tilde \OO_1(x) \bar {\tilde\OO}_1(y) \rangle_{\rm Born}$
at order $g^2$.

\subsubsection*{The Representation [2,2,2]}
\label{section:222:details}

Here we have a total of six operators, one of which was missed in
\cite{Ryzhov:2001bp}. The lowest components of superfields
(\ref{eq:222:fund}) are
\begin  {eqnarray}
\label{eq:222:ops} \OO_1 &\equiv& 3 \, \tr z_1 z_1 z_1 z_1 \; \tr
z_2 z_2 - 6 \, \tr z_1 z_1 z_1 z_2 \; \tr z_1 z_2 + \left( 2 \,
\tr z_1 z_1 z_2 z_2 + \tr z_1 z_2 z_1 z_2 \right) \, \tr z_1 z_1
\no \\
\OO_2 &\equiv& \tr z_1 z_1 z_1  ~\tr z_1 z_2 z_2 - \tr z_1 z_1 z_2
~\tr z_1 z_1 z_2
\no \\
\OO_3 &\equiv& \tr z_1 z_1 \left( \tr z_1 z_1 ~\tr z_2 z_2 - \tr
z_1 z_2 ~\tr z_1 z_2 \right)
\no \\
\OO_4 &\equiv& \tr z_1 z_1 z_1 z_1 z_2 z_2 - 2 \, \tr z_1 z_1 z_1
z_2 z_1 z_2 + \tr z_1 z_1 z_2 z_1 z_1 z_2
\no \\
\OO_5 &\equiv& \tr z_1 z_1 z_1 z_2 z_1 z_2 - \tr z_1 z_1 z_2 z_1
z_1 z_2
\no \\
\OO_6 &\equiv&
\tr z_1 z_1 \left( \tr z_1 z_1 z_2 z_2 - \tr z_1 z_2 z_1 z_2 \right)
\end    {eqnarray}
(and we didn't bother to keep the same normalization factors for
all of them --- just whatever looks better). The descendants arise
from the Konishi-like long primary operators as \bea \OO_4 &\sim&
(Q^2 \bar Q^2 )
\; \tr \left[ z_1 z_1 z_j \bar z^j + z_1 z_1 \bar z^j z_j \right]
\no \\
\OO_5 &\sim& (Q^2 \bar Q^2 )
\; \tr \left[ z_1 z_j z_1 \bar z^j \right]
\no \\
\OO_6 &\sim& (Q^2 \bar Q^2 )
\; \left[ \tr z_1 z_1 \right] \; \left[ \tr z_j \bar z^j \right]
\label{eq:konishi:222} \eea
Note that another operator exists in a
long multiplet, whose descendant coincides with $\OO_6$,
\bea
\OO_6 \sim (Q^2 \bar Q^2 )
\; \left[ \tr z_1 z_j \right] \; \left[ \tr z_1 \bar z^j \right]
\eea This may be established by observing
that the difference operator, \bea \label{eq:bianchi:protected} 3
\left[ \tr z_1 z_j \right] \; \left[ \tr z_1 \bar z^j \right] -
\left[ \tr z_1 z_1 \right] \; \left[ \tr z_j \bar z^j \right]
\end{eqnarray}
is semi-short\footnote{This is the non-renormalised 20' operator 
discussed in~\cite{20',Bianchi:2002rw}}.
A similar phenomenon occurs for higher representations,
and we will not mention it explicitly.

\medskip

The linear combinations orthogonal to these operators
at Born level can be taken
as \bea \label{eq:bps:222} \tilde \OO_1 &=& \OO_1 - {24 \over N}
\OO_4 - {48 (2 N^2 - 3) \over N (3 N^2 - 2)} \OO_5 + {40 \over 3
N^2 - 2} \OO_6
\no \\
\tilde \OO_2 &=& \OO_2 - {4 \over N} \OO_4 - {3 (7 N^2 - 8) \over
N (3 N^2 - 2)} \OO_5 + {5 \over 3 N^2 - 2} \OO_6
\no \\
\tilde \OO_3 &=& \OO_3 \hspace{3.6em} - {20 \over 3 N^2 - 2} \OO_5
- {10 \over 3 N^2 - 2} \OO_6 \eea The matrix of two point
functions in this basis is \bea \label{eq:2-pt:222} \left(\matrix{
\langle \tilde\OO_i \tilde\OO_j^\dagger \rangle_{\rm Born} & 0 \cr
0 & \langle L_i L_j^\dagger \rangle_{\rm Born} }\right) +
\left(\matrix{ 0 & 0 \cr 0 & \langle L_i L_j^\dagger \rangle_{g^2}
}\right) \eea where the (symmetric) blocks are \bea
\label{eq:222:components} \langle L_i L_j^\dagger \rangle_{\rm
Born} = \mbox{$ {(N^2 - 1) \over 64} $} \left(\matrix{ 7 N^4 + 20
N^2 + 8 & -4 N^4 -10 N^2 +4 & 2 N (13 N^2 - 2) \cr
                   & 3 N^4 + 2              & - 2 N (6 N^2 + 1) \cr
                   &                        & 2 N^2 (3 N^2 + 13)
}\right) \eea and \bea \langle L_i L_j^\dagger \rangle_{g^2} =
\mbox{$ {3 \tilde B N (N^2 - 1) \over 32} $} \left(\matrix{ 25 N^4
+ 148 N^2 + 8 & 15 N^4 - 66 N^2 + 4    & 4 N (27 N^2 + 17) \cr
                     & 10 N^4 + 22 N^2 + 2 & - 2 N (28 N^2 + 13) \cr
                     &                        & 2 N^2 (9 N^2 + 89)
}\right)
\no \\
\eea and \bea \langle \tilde\OO_1 \tilde\OO_1^\dagger \rangle_{\rm
Born} & = & {360 ~\cC_N \over N^2 (3 N^2 - 2)} \times
 (N^6 - 11 N^4 + 70 N^2 - 48)
\no \\
\langle \tilde\OO_1 \tilde\OO_2^\dagger \rangle_{\rm Born} & = &
{-120 ~\cC_N \over N^2 (3 N^2 - 2)} \times
 (5 N^4 - 36 N^2 + 24)
\no \\
\langle \tilde\OO_1 \tilde\OO_3^\dagger \rangle_{\rm Born} & = &
{240 ~\cC_N \over N (3 N^2 - 2)} \times
 (N^2 - 2) (2 N^2 - 3)
\no \\
\langle \tilde\OO_2 \tilde\OO_2^\dagger \rangle_{\rm Born} & = &
{5 ~\cC_N \over N^2 (3 N^2 - 2)} \times (3 N^6 - 41 N^4 + 160 N^2
- 96)
\no \\
\langle \tilde\OO_2 \tilde\OO_3^\dagger \rangle_{\rm Born} & = &
{-20 ~\cC_N \over N (3 N^2 - 2)} \times
  (13 N^2 - 12)
\no \\
\langle \tilde\OO_3 \tilde\OO_3^\dagger \rangle_{\rm Born} & = &
{60 ~\cC_N \over  (3 N^2 - 2)} \times
 (N^2 + 1) (N^2 - 2)
.
\eea
Here and below we shall use the abbreviation,
\bea
\label{eq:constant:cN:defined} \cC_N  \equiv
(N^2 -1 ) (N^2 -4) / 64
.
\eea

As seen from (\ref{eq:2-pt:222}), the operators defined in
(\ref{eq:bps:222}) have protected two-point functions $\langle
\tilde\OO_i \tilde\OO_j^\dagger \rangle$ at order $g^2$. This
shows that we can argue that $\tilde \OO_1$, $\tilde \OO_2$,
$\tilde \OO_3$ are the 1/4-BPS primaries we are after. Anomalous
scaling dimensions of long operators ($L_i = \OO_4, \OO_5, \OO_6$)
match those of their Konishi-like primaries computed in
\cite{Bianchi:2002rw}.

\subsubsection*{A Better basis for protected [2,2,2] operators}

It may seem odd that the operators mixing of the operators in the
representations $[2,0,2]$ and $[2,1,2]$ are in terms of
coefficients that are merely inverse powers of $N$, while the
mixing coefficients for the operators we identified in the
representation $[2,2,2]$ have more complicated denominators. This
distinction would also be surprising from the perspective of
AdS/CFT, since the more complicated denominators would suggest
that an infinite series of corrections in the string coupling $g_s
= \lambda /N$ would appear for given `t Hooft coupling $\lambda$.
As a matter of fact, the mixing coefficients depend upon the bases
chosen for both the $\OO_1$, $\OO_2$ and $\OO_3$ operators as well
as the pure descendants. In a different basis, the coefficients
are all proportional to inverse powers of $N$. For the
representation $[2,2,2]$, these new operators are found easily,
and we have \bea
\tilde \OO_1' &=& \tilde \OO_1 + {4 \over N} \tilde \OO_3 = \OO_1
+ {4 \over N} \OO_3 - {24 \over N} \OO_4 - {32 \over N} \OO_5
\no \\
\tilde \OO_2' &=& \tilde \OO_2 + {1 \over 2 N} \tilde \OO_3 =
\OO_2 - {4 \over N} \OO_4 - {7 \over N} \OO_5
\no \\
\tilde \OO_3' &=& \tilde \OO_3 + {2 \over 3 N} \tilde \OO_1 - {4
\over N} \tilde \OO_2 = \OO_3  + {2 \over 3 N} \OO_1 - {4 \over N}
\OO_2 - {10 \over 3 N} \OO_6 \label{eq:bps:222:better} \eea In
this new basis, the matrix of 2-pt functions now reads \bea
\label{eq:222:components:better} \langle \tilde\OO_i'
\tilde\OO_j'{}^\dagger \rangle_{\rm Born} &=&
{5 \cC_N \over N^2} \left(\matrix{ 24 (N^4 + 3 N^2 + 32) & -16
(N^2 + 9)       & 16 N (4 N^2 - 1) \cr
                      & (N^2 - 3) (N^2 + 9) & - 2 N (N^2 - 9)  \cr
                      &                     & 4 (N^4 + 17 N^2 - 24)
}\right)
\no \\
\eea

\subsubsection*{The Representation [2,3,2]}
\label{section:232:details}

Here there is a total of 8 operators (one was overlooked in
\cite{Ryzhov:2001bp}). The operators corresponding to the basis
(\ref{ops:232}) are \bea \OO_1 &\equiv& 2 \, \tr z_1 z_1 z_1 z_1
z_1 \; \tr z_2 z_2 - 4 \, \tr z_1 z_1 z_1 z_1 z_2 \; \tr z_1 z_2
\no \\ && \hskip .5in + \left( \tr z_1 z_1 z_1 z_2 z_2 + \tr z_1
z_1 z_2 z_1 z_2 \right) \tr z_1 z_1
\no \\
\OO_2 &\equiv& 3 \, \tr z_1 z_1 z_1 z_1 \; \tr z_1 z_2 z_2 - 6 \,
\tr z_1 z_1 z_1 z_2 \; \tr z_1 z_1 z_2 \no \\ && \hskip .5in +
\left( 2 \, \tr z_1 z_1 z_2 z_2 + \tr z_1 z_2 z_1 z_2 \right) \tr
z_1 z_1 z_1 \quad\quad
\no \\
\OO_3 &\equiv& \tr z_1 z_1 z_1 \left( \tr z_1 z_1 \; \tr z_2 z_2 -
\tr z_1 z_2 \; \tr z_1 z_2 \right)
\no \\
\OO_4 &\equiv& \tr z_1 z_1 z_1 \; \tr z_1 z_1 \; \tr z_2 z_2 -2 \,
\tr z_1 z_1 z_2 \; \tr z_1 z_1 \; \tr z_1 z_2
+ \tr z_1 z_2 z_2 \; \tr z_1 z_1 \; \tr z_1 z_1
\no \\
\OO_5 &\equiv& \tr z_1 z_1 z_1 z_1 z_1 z_2 z_2 - 2 \, \tr z_1 z_1
z_1 z_1 z_2 z_1 z_2 + \tr z_1 z_1 z_1 z_2 z_1 z_1 z_2
\no \\
\OO_6 &\equiv& \tr z_1 z_1 z_1 z_1 z_2 z_1 z_2 - \tr z_1 z_1 z_1
z_2 z_1 z_1 z_2
\no \\
\OO_7 &\equiv& \left( \tr z_1 z_1 z_1 z_2 z_2 - \tr z_1 z_1 z_2
z_1 z_2 \right) \; \tr z_1 z_1
\no \\
\OO_8 &\equiv& \left( \tr z_1 z_1 z_2 z_2 - \tr z_1 z_2 z_1 z_2
\right) \; \tr z_1 z_1 z_1 \label{2-3-2:operators} \eea Out of
these, four are descendants, \bea \OO_5 &\sim& (Q^2 \bar Q^2 )
\; \tr \left[ 2 z_1 z_1 z_1 z_j \bar z^j + 2 z_1 z_1 z_1 \bar z^j
z_j - z_1 z_1 z_j z_1 \bar z^j - z_1 z_1 \bar z^j z_1 z_j \right]
\no \\
\OO_6 &\sim& (Q^2 \bar Q^2 )
\; \tr \left[ z_1 z_1 z_1 z_j \bar z^j + z_1 z_1 z_1 \bar z^j z_j
- 2 z_1 z_1 z_j z_1 \bar z^j - 2 z_1 z_1 \bar z^j z_1 z_j \right]
\no \\
\OO_7 &\sim& (Q^2 \bar Q^2 )
\left[ \tr z_1 z_1 z_j \; \tr z_1 \bar z^j + \tr z_1 z_1 \bar z^j
\; \tr z_1 z_j - 12 \, \tr z_1 z_1 z_1 \; \tr z_j \bar z^j \right]
\no \\
\OO_8 &\sim& (Q^2 \bar Q^2 )
\left[ \tr z_1 z_1 z_1 \; \tr z_j \bar z^j \right]
\label{eq:konishi:232} \eea while the combinations orthogonal to
them at Born level
can be taken as \bea \label{eq:bps:232} \tilde \OO_1 &=&
\OO_1 - {20 \over N} \OO_5 - {30 (N^2 - 2) \over N^3} \OO_6 + {15
(N^2 - 2) \over N^4} \OO_7 + {10 (N^2 + 2) \over N^4} \OO_8
\no \\
\tilde \OO_2 &=& \OO_2 - {30 \over N} \OO_5 - {30 (2 N^2 - 3)
\over N^3} \OO_6 + {15 (2 N^2 - 3) \over N^4} \OO_7 + {10 (N^2 +
3) \over N^4} \OO_8
\no \\
\tilde \OO_3 &=& \OO_3 \hspace{3.6em} - {12 \over N^2} \OO_6 - {3
(N^2 - 2) \over N^3} \OO_7 - {2 (N^2 + 2) \over N^3} \OO_8
\no \\
\tilde \OO_4 &=& \OO_4 \hspace{3.6em} - {18 \over N^2} \OO_6 - {(7
N^2 - 9) \over N^3} \OO_7 - {2 (2 N^2 + 9) \over 3 N^3} \OO_8 \eea
The matrix of two point functions in this basis is \bea
\label{eq:2-pt:232} \left(\matrix{ \langle \tilde\OO_i
\tilde\OO_j^\dagger \rangle_{\rm Born} & 0 \cr 0 & \langle L_i
L_j^\dagger \rangle_{\rm Born} }\right) + \left(\matrix{ 0 & 0 \cr
0 & \langle L_i L_j^\dagger \rangle_{g^2} }\right) \eea so indeed
at order $g^2$ the operators defined in (\ref{eq:bps:232}) have
protected correlators. This shows that we can argue that $\tilde
\OO_1$, $\tilde \OO_2$, $\tilde \OO_3$, $\tilde \OO_4$ are the
1/4-BPS primaries we are after. The (symmetric) blocks in equation
(\ref{eq:2-pt:232}) are \bea \label{eq:232:components:begin}
\langle L_i L_j^\dagger \rangle_{\rm Born} = \mbox{$ \half N \cC_N
$} \left(\matrix{ 6 N^2 + 45 & -3 N^2 - 9 & 18 N         & 36 N
\cr
               & 2 N^2 - 3    & -4N          & - 15 N \cr
               &              &  4 N^2 + 24 & 36     \cr
               &              &              & 9 N^2 + 54
}\right) \eea and \bea \langle L_i L_j^\dagger \rangle_{g^2} & = &
\mbox{$ 12 \tilde B N \cC_N $} \no \\ &&\times \left(\matrix{ 9 N
(2 N^2 + 27) & - 9 N (N^2 + 8) & 54 (N^2 + 2)   & 27 (5 N^2 + 6)
\cr
                 & N (5 N^2 + 17)  & -6 (3 N^2 + 2)  & -6 (10 N^2 + 3) \cr
                 &                  & 4 N (2 N^2 + 23) & 174 N         \cr
                 &                  &                  & 9 N (3 N^2 + 35)
}\right)
\no \\
\label{eq:232:components} \eea The Born level overlaps of
protected operators are given by ugly and not particularly
illuminating expressions. Here we list them for the sake of
completeness: \bea \label{eq:232:protected} \langle \tilde\OO_1
\tilde\OO_1^\dagger \rangle_{\rm Born} &=&
 30 ~\cC_N N^{-5} (N^8 - 10 N^6 + 117 N^4 - 720 N^2 + 420)
\no \\
\langle \tilde\OO_1 \tilde\OO_2^\dagger \rangle_{\rm Born} &=& -90
~\cC_N N^{-5} (4 N^6 - 69 N^4 + 395 N^2 - 210)
\no \\
\langle \tilde\OO_1 \tilde\OO_3^\dagger \rangle_{\rm Born} &=&
 90 ~ \cC_N N^{-4} (N^2 - 2) (N^4 - 7 N^2 + 14)
\no \\
\langle \tilde\OO_1 \tilde\OO_4^\dagger \rangle_{\rm Born} &=&
 30 ~ \cC_N N^{-4} (N^2 - 2) (N^2 - 9) (3 N^2 - 7)
\no \\
\langle \tilde\OO_2 \tilde\OO_2^\dagger \rangle_{\rm Born} &=&
 45 ~ \cC_N
 N^{-5} (N^8 - 29 N^6 + 328 N^4 - 1290 N^2 + 630)
\no \\
\langle \tilde\OO_2 \tilde\OO_3^\dagger \rangle_{\rm Born} &=&
-630 ~\cC_N  N^{-4} (N^2 - 1) (N^2 - 6)
\no \\
\langle \tilde\OO_2 \tilde\OO_4^\dagger \rangle_{\rm Born} &=&
 30 ~ \cC_N
 N^{-4} (N^2 - 1) (N^2 - 9) (2 N^2 - 21)
\no \\
\langle \tilde\OO_3 \tilde\OO_3^\dagger \rangle_{\rm Born} &=&
 9 ~ \cC_N N^{-3} (N^6 - N^4 - 16 N^2 + 56)
\no \\
\langle \tilde\OO_3 \tilde\OO_4^\dagger \rangle_{\rm Born} &=&
 6 ~ \cC_N N^{-3} (N^2 - 9) (N^4 + 3 N^2 - 14)
\no \\
\langle \tilde\OO_4 \tilde\OO_4^\dagger \rangle_{\rm Born} &=&
 2 ~\cC_N N^{-3} (N^2 - 9) (7 N^4 + 16 N^2 - 63)
\eea and the constant $\cC_N = (N^2 - 1) (N^2 - 4) / 64$ was
defined in (\ref{eq:constant:cN:defined}).

\subsubsection*{The Representation [3,1,3]}
\label{section:313:details}

Here there is a total of 5 operators (one was overlooked in
\cite{Ryzhov:2001bp}). The highest weight states lowest component
operators corresponding to the basis (\ref{ops:313}) are \bea
\OO_1 &\equiv& \tr z_1 z_1 z_1 z_1 ~\tr z_2 z_2 z_2 - 3 \, \tr z_1
z_1 z_1 z_2 ~\tr z_1 z_2 z_2 \no \\ && \hskip .5in + \left( 2 \,
\tr z_1 z_1 z_2 z_2 + \tr z_1 z_2 z_1 z_2 \right) \tr z_1 z_1 z_2
- \tr z_1 z_2 z_2 z_2 ~\tr z_1 z_1 z_1
\nonumber\\
\OO_2 &\equiv& \tr z_1 z_2 \left( 2 \, \tr z_1 z_2 ~\tr z_1 z_2
z_2 - \tr z_2 z_2 ~\tr z_1 z_1z_1 - 3 \, \tr z_1 z_1 ~\tr z_1
z_2z_2 \right) \no \\ && \hskip .5in + \tr z_1 z_1 \left( \tr z_2
z_2 ~\tr z_1 z_1 z_2 + \tr z_1 z_1 ~\tr z_2 z_2z_2 \right)
\no \\
\OO_3 &\equiv& \tr z_1 z_1 z_1 z_2 z_1 z_2 z_2 - \tr z_1 z_1 z_1
z_2 z_2 z_1 z_2
\no \\
\OO_4 &\equiv& 2 \, \tr z_1 z_1 z_1 z_1 z_2 z_2 z_2 - 3 \, \tr z_1
z_1 z_1 z_2 z_1 z_2 z_2 - 3 \, \tr z_1 z_1 z_1 z_2 z_2 z_1 z_2 \no
\\ && \hskip .5in + 2 \, \tr z_1 z_1 z_2 z_1 z_2 z_1 z_2 + 2 \,
\tr z_1 z_1 z_2 z_2 z_1 z_1 z_2
\no \\
\OO_5 &\equiv& - \left( \tr z_1 z_1 z_2 z_2 z_2 - \tr z_1 z_2 z_1
z_2 z_2 \right) \tr z_1 z_1 \no \\ && \hskip .5in + \left( \tr z_1
z_1 z_1 z_2 z_2 - \tr z_1 z_1 z_2 z_1 z_2 \right) \tr z_1 z_2
\end    {eqnarray}
Out of these three are descendants, \bea \OO_3 &\sim& (Q^2 \bar
Q^2 )
\; \tr \left[ z_1 z_1 z_2 z_j \bar z^j + z_1 z_1 z_2 \bar z^j z_j
- z_1 z_1 z_j \bar z^j z_2 - z_1 z_1 \bar z^j z_j z_2 \right]
\no \\
\OO_4 &\sim& (Q^2 \bar Q^2 )
\; \tr \left[ z_1 z_1 z_2 z_j \bar z^j + z_1 z_1 z_2 \bar z^j z_j
+ z_1 z_1 z_j \bar z^j z_2 + z_1 z_1 \bar z^j z_j z_2
\right.\nonumber\\&&\left.\hspace{6em} - 2 z_1 z_2 z_1 z_j \bar
z^j - 2 z_1 z_2 z_1 \bar z^j z_j
\right] \\
\OO_5 &\sim& (Q^2 \bar Q^2 )  \; \left[ 2 \, \tr z_1 z_1 z_j \;
\tr z_2 \bar z^j + 2 \, \tr z_1 z_1 \bar z^j \; \tr z_2 z_j - \tr
z_1 z_2 \bar z^j \; \tr z_1 z_j
\right.\nonumber\\&&\left.\hspace{4em} - \tr z_1 z_2 z_j \; \tr
z_1 \bar z^j - \tr z_2 z_1 \bar z^j \; \tr z_1 z_j - \tr z_2 z_1
z_j \; \tr z_1 \bar z^j \right] \quad\quad \label{eq:konishi:313}
\no \eea while the combinations orthogonal to them at Born level
can be taken as
\bea \label{eq:bps:313} \tilde \OO_1 &=& \OO_1 - {2 N \over N^2 -
2} \OO_4 - {5 \over N^2 - 2} \OO_5
\no \\
\tilde \OO_2 &=& \OO_2 + {8 \over N^2 - 2} \OO_4 + {10 N \over N^2
- 2} \OO_5 \eea The matrix of two point functions in this basis is
\bea \label{eq:2-pt:313} \left(\matrix{ \langle \tilde\OO_i
\tilde\OO_j^\dagger \rangle_{\rm Born} & 0 \cr 0 & \langle L_i
L_j^\dagger \rangle_{\rm Born} }\right) + \left(\matrix{ 0 & 0 \cr
0 & \langle L_i L_j^\dagger \rangle_{g^2} }\right) \eea so indeed
at order $g^2$ the operators defined in (\ref{eq:bps:313}) have
protected correlators. This shows that we can argue that $\tilde
\OO_1$, $\tilde \OO_2$, are the 1/4-BPS primaries we are after.
The (symmetric) blocks in (\ref{eq:2-pt:313}) are \bea
\label{eq:313:components} \langle L_i L_j^\dagger \rangle_{\rm
Born} &=& \mbox{$
 \cC_N
$} \left(\matrix{ N (N^2 - 9) & 0              & 0        \cr
              & 15 N (N^2 + 3) & - 30 N^2 \cr
              &                & 3 N (N^2 + 6)
}\right)
\no \\
\langle L_i L_j^\dagger \rangle_{g^2} &=& \mbox{$ 6 \tilde B N
\cC_N $} \left(\matrix{ 5 N (N^4 - 9) & 0              & 0
\cr
                & 75 N (N^2 + 7) & 180 (N^2 + 1) \cr
                &                & 12 N (N^2 + 16)
}\right)
\no \\
\langle \tilde\OO_i \tilde\OO_j^\dagger \rangle_{\rm Born} &=& {15
(N^2 - 9) \over  N^2 (N^2 - 2)} ~ \cC_N \left(\matrix{ (N^2 - 1)
(N^2 - 4) & 4 N (N^2 - 1) \cr
                    & 2 N^2 (N^2 - 6)
}\right) \eea

Here we should mention that the operator $\OO_3$ (which was
overlooked in \cite{Ryzhov:2001bp}) has zero correlators with
everything else. The reason is that $\cO_3^\dagger = - (\cO_3)^*$
while all other operators satisfy $\cO_i^\dagger = + (\cO_i)^*$.

\subsubsection*{Completeness of the Construction}

An important point which remains to be addressed is whether the
construction of the 1/4 BPS operators given above is exhaustive.
The fact that it is follows from $SU(4)$ group theory in the
following manner. Given a 1/4 BPS representation of $SU(4)$ of the
type $[q,p,q]$, one begins by listing all possible monomial scalar
composite operators built out of $(p+q)$ $z_1$'s
and $q$ $z_2$'s.%
\footnote{
    The remaining scalar field $z_3$ never enters
    in the highest weight of a 1/4 BPS representation, though it will
    also be needed when describing 1/8 BPS operators.
} These monomials form a basis for the linear space of scalar
composite operators of the form $[ (z_1)^{(p+q)} \, (z_2)^q ]$.
They can occur in representations $[0,p+2q,0]$, $[1,p+2q-2,1]$,
... , $[q,p,q]$. Then we have to show that the number of such
monomials matches the total number of operators we constructed in
these representations.

Let us illustrate how this works with an example. Consider the
[2,2,2] representation. The complete set of scalar composite
operators we can build out of 4 $z_1$'s and 2 $z_2$'s is
\begin  {eqnarray}
\label{eq:222:monomials} 6: && \tr z_1 z_1 z_1 z_1 z_2 z_2 , ~~
\tr z_1 z_1 z_1 z_2 z_1 z_2 , ~~ \tr z_1 z_1 z_2 z_1 z_1 z_2
\no \\
4+2: && \tr z_1 z_1 z_1 z_1 \; \tr z_2 z_2 , ~~ \tr z_1 z_1 z_1
z_2 \; \tr z_1 z_2 , ~~ \tr z_1 z_1 z_2 z_2 \; \tr z_1 z_1 , ~~
\tr z_1 z_2 z_1 z_2 \; \tr z_1 z_1
\no \\
3+3: && \tr z_1 z_1 z_1 \; \tr z_1 z_2 z_2 , ~~ \tr z_1 z_1 z_2 \;
\tr z_1 z_1 z_2
\no \\
2+2+2: && \tr z_1 z_1 \; \tr z_1 z_1 \; \tr z_2 z_2 , ~~ \tr z_1
z_1 \; \tr z_1 z_2 \; \tr z_1 z_2
\end    {eqnarray}
or the total of 11 operators. By taking linear combinations of
these, we can construct:

$\bullet$ 4 totally symmetric tensors in the [0,6,0] corresponding
to the partitions of 6 in (\ref{eq:222:monomials});

$\bullet$ 1 tensor in the representation [1,4,1], given in
(\ref{eq:141:ops});

$\bullet$ 6 tensors in the representation [2,2,2], listed in
(\ref{eq:222:ops}).

Thus there are no other scalar composite operators of the form $[
(z_1)^4 \, (z_2)^2 ]$.

\medskip

In the same fashion, we can go through all other representations
we have considered in this paper and verify that we didn't leave
out any operators.


\end{document}